\documentstyle[12pt,epsfig]{article}

\newcommand{\be}{\begin{equation}}
\newcommand{\ee}{\end{equation}}
\begin{document}  
\topmargin 0pt
\oddsidemargin=-0.4truecm
\evensidemargin=-0.4truecm
\renewcommand{\thefootnote}{\fnsymbol{footnote}}
\newpage
\setcounter{page}{0}
\begin{titlepage}   
\vspace*{-2.0cm}  
\begin{flushright}
hep-ph/yymmnn
\end{flushright}
\vspace*{0.1cm}
\begin{center}
{\Large \bf Model Independent Constraints on Solar Neutrinos} \\ 
\vspace{0.6cm}

\vspace{0.4cm}

{\large 
Lal Singh$^a$\footnote{lalsingh96@yahoo.com},
Bhag C. Chauhan$^b$\footnote{bcawake@gmail.com} 
Ravi Dutt $^c$\footnote{ravidutt18@yahoo.com}
K. K. Sharma$^c$\footnote{kks@nith.ac.in} and 
S. Dev$^a$\footnote{dev5703@yahoo.com}\\
}
\vspace{0.15cm}
$^a$ {\small \sl Department of Physics \\
Himachal Pradesh University, Shimla (HP) INDIA-171005. }\\
$^b$ {\small \sl Department of Physics\\
Govt. College Karsog, Distt. Mandi (HP) INDIA-171304.} \\
$^c$ {\small \sl Department of Physics \\
National Institute of Technology, Hamirpur (HP) INDIA-177005.}\\
\end{center}
\vglue 0.6truecm

\begin{abstract}
Using the data from SNO NCD phase, SuperK, Borexino and KamLAND Solar phase, we derive in a model independent way, bounds on the possible components in the solar neutrino flux. 
We update the limits on the antineutrino ($\bar\nu_x$) flux and sterile ($\nu_s$) component and compare them with the previous results obtained using SNO Salt phase data and data from SuperKamiokande experiments. It is affirmed that the upper bound on $\bar\nu_x$ is independent of the $\nu_s$ component. We recover the $\nu_s$ and $\bar\nu_x$ upper bounds existing in the literature. We also obtain bounds on $f_B$, the SSM normalization factor and the common parameter range for $f_B$ and the $\nu_s$ components in the light of latest data. In summary, we update, in a model independent way, the previous results existing in literature in the light of latest solar neutrino data.   
\end{abstract}
\end{titlepage}   

\renewcommand{\thefootnote}{\arabic{footnote}}

\setcounter{footnote}{0}

\section{Introduction}
The history of solar neutrino experiments begins in the early 1960's with the
Homestake Solar neutrino detector and its prototypes. The purpose of this experiment was to verify the fusion reactions that power the Sun by measuring the resulting neutrino flux. Instead of confirming the predictions of the Standard Solar Model (SSM) it measured a significant deficit which came to be known as the Solar Neutrino Problem (SNP) \cite{Davis:1968cp, Cleve:1998}. 

The mystery of the missing neutrinos deepened as subsequent experiments were performed. After a journey of about four decades we are standing on a square where we have in hand a leading solution for the SNP. The KamLAND experiment \cite{Eguchi:2002dm} has acknowledged neutrino oscillations through Large Mixing Angle (LMA) \cite{LMA} as the dominant solution for the solar neutrino deficit \cite{Davis:1968cp}. It has become evident that the mechanism of Spin Flavour Precession (SFP) to active antineutrinos in the Sun is either absent or plays a subdominant role. In fact these active neutrinos would originate a sizable $\bar\nu_e$ flux, whose upper bound has become stricter and corresponds to 0.028\% of the $^8 B$ neutrino flux \cite{Eguchi:2003gg}. 

The analysis of the available solar neutrino data done by Peter Sturrock et al. \cite{Sturrock:2004hx} have on the other hand provided increasing evidence that the neutrino flux from the Sun is not constant but varies with well-known solar rotation periods. If such findings are confirmed ever in future, the need for an addition to the LMA solution will be obvious and will most likely rely on an interaction of the solar magnetic field with the neutrino  magnetic moment. Since an SFP conversion to active antineutrinos is unlikely, this interaction is expected to produce a significant and time varying flux of sterile neutrinos \cite{Berezinsky:2002fa}, \cite{deHolanda:2002ma}, \cite{Chauhan:2004sf}. 

At present, we have solar neutrino data from several neutrino experiments including the Neutral Current Detectors (NCDs) phase of Sudbury Neurtino Observatory (SNO), SuperK-III, Borexino and KamLAND Solar phase \cite{ncd, sk, Bor, kl}. In this paper, we have performed a model independent analysis of the latest available solar neutrino data and derived constraints on the sterile neutrino flux and active antineutrino flux which may accompany the LMA effect. We have, also, derived the corresponding constraints for the normalization factor, $f_B$. Various such analysis \cite{Barger:2001pf, bmw, kang, bcc, skumar} have been done in the past, but in this work we are corroborating and further constraining the previous findings in the light of recent experimental data. 
In section 2 we details the data available from all the solar neutrino experiments. In section 3 we present the theory of the model independent analysis. We start with the three  master equations of neutrino flux and derived several other equations relevant for our predictions. We also use another version of the three master equations as neutrino flux rates for constraining the $f_{B}$. $\chi^{2}$-Fitting is also presented in this section. In section 4 we discuss our results and the conclusions are finally summarized in the last section.  
\section{Solar Neutrino Data}
Here we discuss the only solar neutrino experiments which are relevant for our present study. In our analysis we use data from SNO, SuperK, Borexino, Homestake and KamLAND experiments.
\subsection{SNO and SuperK}
The Sudbury Neutrino Observatory (SNO) detects $^{8}B$ solar neutrinos through three reactions: charged-current interactions(CC) on deuteron, in which only electron neutrinos participate; neutrino-electron elastic scattering (ES), which are dominated by contributions from electron neutrinos and neutral-current (NC) disintegration of the deuteron by neutrinos, which has equal sensitivity to all active neutrino flavors.
The SNO experiment had three stages of running. The first stage was pure $D_{2}O$ from November 1999 to May 2001. The second stage or SNO-II (Salt phase) from June 2001 to October 2003 where 2000 kg of NaCl was added to the $D_{2}O$ to increase the neutron detection efficiency. The third and final phase saw the removal of the salt and the addition 36 strings of $^{3}He$ proportional counters, Neutral Current Detectors (NCDs), to provide an independent detection of neutrons. This phase also known as SNO-III (NCDs)  phase, ran from November 2004 to November 2006. The three stages of the SNO running can be thought of three distinct experiments measuring the flux of $^{8}B$ solar neutrinos flux with the neutral current reaction as these three stages have very different systematic uncertainties for the detection of neutrons. 

The second phase of running  for a data set of 254.2 days and a 5.5 MeV energy threshold and  the third phase of running with 385.2 days of data and a 5 MeV energy threshold following the installation of the NCDs measure the three fluxes as shown in Table 1.

Super-Kamiokande is a cylindrical 50 kt. water Cerenkov detector which observes high energy solar neutrinos via elastic scattering of electrons. The Super-Kamiokande experiment started taking data in April, 1996 and continued the observation for five years within the running period referred to SK-I till the detector maintenance in July, 2001. The Super-Kamiokande detector was rebuilt after the accident with the half of the original PMT density in the inner detector and resumed observation from October, 2002, which is referred to the SK-II running period.
The SK-II continued the physics measurement for three years and finished in October 2005 for the reconstruction work to put the PMT density back to the SK-I level. The Super-Kamiokande detector has restarted observation in June, 2006, which is referred to the SK-III period. New electronics was installed on the detector in September 2008, starting the SK-IV running period.

In this paper, the data observed in the SK-II (2002-2005) and SK-III (2006-2008) running periods are used. 
The SK-II and SK-III flux measured in the experiment is shown along with SNO data in the Table 1.
In Table 2 we present the data of Table 1 in terms of the corresponding rates with reference to the SSM flux \cite{SSM}.
\begin{table}
\begin{small}
\begin{center}
\begin{tabular}{|c|c|c|c|}  \hline  
        &        $\phi^{CC}$         &    $ \phi^{NC}$         &  $\phi^{ES}$        \\  \hline 
SNO-II  &  $1.68 ^{+}_{-} 0.11$     &  $4.94 ^{+}_{-} 0.43 $  &  $2.35 ^{+}_{-} 0.27$        \\ 
SNO-III &  $1.67 ^{+}_{-} 0.09$      & $5.54 ^{+}_{-} 0.49$    & $1.77 ^{+}_{-} 0.26$       \\ 
SK-II   &                 -           &      -                   &  $2.38 ^{+}_{-} 0.17$        \\ 
SK-III  &                 -           &       -                  & $2.32 ^{+}_{-} 0.06$       \\  \hline 
\end{tabular}
\end{center}
\caption{Solar Neutrino flux measured at SNO \cite{sno} \& SK \cite{sk} in units of $10^{6}\ cm^{-2}s^{-1}$.}
\end{small}
\end{table}
\begin{table}
\begin{small}
\begin{center}
\begin{tabular}{|c|c|c|c|}  \hline  
        &       $ R^{CC}$                  &   $R^{NC}$             &  $R^{ES}$        \\  \hline 
SNO-II  &  $0.286^{+}_{-} 0.02$            &  $0.840^{+}_{-} 0.07$  &  $0.400^{+}_{-} 0.05$        \\ 
SNO-III & $0.284^{+}_{-} 0.02$             & $0.942^{+}_{-} 0.08$   & $0.301^{+}_{-} 0.04$       \\  
SK-II   &          -          &           -                           &  $0.405^{+}_{-} 0.02$      \\
SK-III  &          -          &           -                          &  $0.395^{+}_{-} 0.01$     \\ \hline 
\end{tabular}
\end{center}
\caption{Different rates with $1\sigma$ errors for the SNO and SK experiments.}
\end{small}
\end{table}
\subsection{Borexino, Homestake and KamLAND}
Borexino is a low threshold liquid scintillator detector for solar neutrinos. Solar neutrinos of medium energy range ($^{7}Be$, CNO, pep) are detected in BOREXINO via elastic scattering of electrons. The detector is located in underground laboratory at Gran Sasso, Italy. Because of the ultra-high radio purity it is the first experiment able to do a real-time analysis of low energy solar neutrinos.  As a target a 300t of liquid scintillator is used. The scintillator is contained in a spherical nylon vessel. Outside a non-scintillating buffer liquid acts as passive shielding. The scintillation light is registered by more than 2200 photomultipliers (PMs) mounted on the inner surface of a stainless steel sphere.  Additional 205 PMs on the outside surface of the sphere and at the floor of the dome are mounted. Hence, the water volume acts as shielding against external gamma and neutron radiation and as an active muon veto. Borexino reports the interaction rate of the $0.862~Mev$ $^{7}Be$ solar neutrino $49~\pm~3_{stat}~\pm~4_{syst}$ counts/(day.100 ton) for 192 live days data in the period from May 16, 2007 to April, 12, 2008 and $P^{M}_{ee}=0.56^{+}_{-}0.10$ ($1\sigma$) \cite{Marco, 0805}.   

The Homestake experiment was one of the longest continuously running physics
experiments. The experiment started taking data in 1967 and released its first results in 1968 \cite{Davis:1968cp}. After several upgrades, data taking resumed in 1970 and the experiment proceeded to collect data almost continuously until 1994. The heart of the Homestake detector was 615 tons of perchloroethylene, $C_{2}Cl_{2}$ or dry cleaning fluid. The neutrinos were detected via the reaction: $^{37}Cl + \nu_{e} -\rightarrow ^{37}Ar + e^{-}$. Homestake predict solar neutrino flux  $2.56^{+}_{-} 0.16$ SNU \cite{Cleve:1998}. 

KamLAND is acronym for the Kamioka Liquid Scintillating Anti-Neutrino Detector. The detector is located in the Kamioka Mine near the city of Kamioka in the Gifu Prefecture of Japan. KamLAND occupies the old Kamiokande site within the mine. The KamLAND detector uses 1000 metric tons of liquid scintillator, abbreviated 1 kilo-ton or 1kt, as both the target and detection medium for low energy nuclear/particle physics processes like neutrino elastic scattering and inverse beta decay. The molecules that compose the liquid scintillator give off light when charged particle move through the detector. KamLAND is designed and instrumented to detect this light and reconstruct the physics processes that produce the light.The data for $^8B$ solar neutrino flux has been observed in the experiment which is not of our interest here \cite{thesiskamLand}. However, the data for $^7Be$ solar neutrino flux is expected to come \cite{7Be}
\section{Theory of Model Independent Analysis}
The model independent equations in terms of solar neutrino flux, neglecting electronic antineutrino component, are given as \cite{skumar}
\begin{center}
\begin{eqnarray}
\phi^{CC}=\phi_{\nu_{e}},\\
\phi^{NC}=\phi_{\nu_{e}}+\phi_{\nu_{x}}+\bar{r}_{d} \phi_{\bar{\nu}_{x}},\\
\phi^{ES}=\phi_{\nu_{e}}+r\phi_{\nu_{x}}+\bar{r_{x}}\phi_{\bar{\nu}_{x}}.
\end{eqnarray}
\end{center}

We have taken care of the fact that the neutral current (NC) is sensitive equally for all neutrino components, whereas the elastic scattering (ES) is more sensitive to electronic neutrino component than the non-electronic ones. 

The quantities $r$, $\bar{r}_{x}$ are the ratios of the NC neutrino and non-electronic 
antineutrino event rates to the NC+CC neutrino event rate, respectively. However, 
$\bar{r}_{d}$ is the ratio of the antineutrino deuteron fission rate to neutrino 
deuteron fission event rate. We have
\be
r=\frac{\int dE_{\nu}\phi(E_{\nu})\int dE_{e}\int dE^{'}_{e}\frac{d\sigma_{NC}}
{dE_e}f(E^{'}_e,E_{e})}{\sigma_{NC}\rightarrow \sigma_{NC+CC}},
\ee
\be
\bar{r}_{x}=\frac{\int dE_{\nu}\phi(E_{\nu})\int dE_{e}\int dE^{'}_{e}\frac{d\bar\sigma_{NC}}
{dE_e}f(E^{'}_e,E_{e})}{\sigma_{NC}\rightarrow \bar\sigma_{NC+CC}},
\ee
\be
\bar r_{d}=\frac{\int dE_{\nu}\phi(E_{\nu})\bar\sigma_{NC}(E_{\nu})}
{\bar\sigma_{NC}\rightarrow \sigma_{NC}}.
\ee
Here $f$ is the energy resolution function and $\sigma$'s are the cross sections.

It may be noted that we neglect the electronic antineutrino solar neutrino flux as evident from KamLand results \cite{Eguchi:2003gg}.
Here the subscript $`x'$ in $\nu_{x}$ and $\bar{\nu_{x}}$ stands for the non-electronic ($\mu/ \tau$) components.

Using equations (1-3), we express the ES flux as 
\be
\phi^{ES}=r\phi^{NC}+(1-r)\phi^{CC}-(r \bar{r}_{d}-\bar{r}_{x})\phi_{\bar{\nu}_{x}} 
\ee

In the absence of antineutrino component we obtain 
\be
\phi_{no \bar{\nu_{x}}}^{ES}=r\phi^{NC}+(1-r)\phi^{CC}. 
\ee

The non-electronic solar antineutrino flux ($\bar{\nu}_{\mu / \tau}$) is determined as
\be
\phi_{\bar{\nu}_{x}}=\frac{(\phi_{no \bar{\nu_{x}}}^{ES}-\phi^{ES})}{({r\bar{r}_{d}-\bar{r}_{x}})}
\ee
where the term in the denominator is positive definite. So the sign of $\phi_{no \bar{\nu_{x}}}^{ES}-\phi^{ES}$ will tell us whether the $\phi_{\bar\nu_x}$ is present or not in the solar neutrino flux.

The active neutrino ($\nu_{e}+\nu_{x}+ \bar{\nu}_{x}$), non-electronic neutrino ($\nu_{x}$) and  sterile neutrino ($\nu_{sterile}$) fluxes are given by

\begin{eqnarray}
\phi_{active}=\frac{[(r-\bar{r}_{x})\phi^{NC}+(1-\bar{r}_{d})((1-r)\phi^{CC}-\phi^{ES})]}{r\bar{r}_{d}-\bar{r}_{x}},\\
\phi_{\nu_{x}}^{NC}=\phi^{NC}-\phi^{CC}, \\  
\phi_{\nu_{x}}^{ES}=\frac{\phi^{ES}_{SK}-\phi^{CC}}{r}.
\end{eqnarray}
If we substract the active neutrino flux from the SSM predictions we get the flux for sterile neutrinos 
\begin{eqnarray}
\phi_{sterile}=\phi_{SSM}^{B}-\phi_{active}
\end{eqnarray}

Equations (11) and (12)present a special case, in which there are transitions to non-electronic neutrino flux
only, the maximum possible non-electronic neutrino flux can be found from NC flux and ES flux measurements.

Now we derive expressions for another independent way of constraining the neutrino components and the $f_B$. 
If $\psi$ is the mixing angle for $\nu_{x},~\bar{\nu}_{x}$ mixing, the ratio of non-electronic solar neutrino flux to total non-electronic solar neutrino/ antineutrino flux may be written as 
\be
\sin^{2}\psi=\frac{\phi_{\nu_{x}}}{\phi_{\nu_{x}}+\phi_{\bar{\nu}_{x}}}
\ee

Using the model independent flux equations, the above equation can be re-written as
\be
\sin^{2}\psi=\frac{\bar{r}_{d}\gamma-\bar{r}_{x}}{r-\bar{r}_{x}-(1-\bar{r}_{d})\gamma},
\ee
where $\gamma=\frac{\phi^{ES}-\phi^{CC}}{\phi^{NC}-\phi^{CC}}$. 
A similar equation has been obtained in \cite{bcc, skumar}.

To study active-sterile admixture let's take $\alpha$ as the mixing angle between the active and sterile neutrinos, then $sin^{2} \alpha$ denotes the fraction of the all the active neutrinos. The sterile neutrino component is therefore proportional to $\cos^2{\alpha}$. 
The fraction of active neutrino flux excluding electronic component, as measured by CC flux, present in the solar neutrino flux can be calculated by the following relation

\begin{equation}
 \sin^{2}\alpha = \frac{\phi_{active}-\phi^{CC}}{\phi_{SSM}-\phi^{CC}}
\end{equation} 

In order to constrain $f_{B}$, we  analyse the data by using the basic model independent equations in terms of rates. As stated before, the parameter $f_{B}$ is the  normalization to the SSM $^8 B$ neutrino flux \cite{SSM}. 

From the model independent flux equations, the expressions for the charged current(CC), neutral current(NC) and elastic scattering(ES) rates are \cite{bcc} given by
\begin{eqnarray}
R^{CC}=f_{B}P_{ee}, \\
R^{NC}=f_{B}P_{ee}+f_{B}(1-P_{ee})[\sin^{2}\alpha\sin^{2}\psi+\bar
{r}_{d}\sin^{2}\alpha\cos^{2}\psi], \\
R^{ES}=f_{B}P_{ee}+f_{B}(1-P_{ee})[r\sin^{2}\alpha\sin^{2}\psi+\bar{
r_{x}}\sin^{2}\alpha\cos^{2}\psi].
\end{eqnarray}

Owing to its near energy independence in this range, the electron neutrino survival 
probability $P_{ee}$ is factored out of these integrals as in eqs.(17)-(19).  
It is evident from the above equations that the electron neutrinos converted into the other flavours are proportional to $1-P_{ee}$.

The normalization to SSM $^{8}B$ neutrino flux can be obtained from equations (17)
and (18) as 
\be 
f_{B}=R^{CC}+\frac{(R^{NC}-R^{CC})}{\sin^{2}\alpha(\sin^{2}\psi+\bar
{r}_{d}\cos^{2}\psi)}
\ee
For no-sterile case ($\sin^{2}\alpha=1$), $f_{B}$ from the above equation becomes
\be
f_{B}=R^{CC}+\frac{(R^{NC}-R^{CC})}{(\sin^{2}\psi+\bar{r}_{d}\cos^{2}\psi)}
\ee
Using equations (17), (18) and (19), for no-sterile case ($\sin^{2}\alpha=1$), the constraints on $f_{B}$ can be obtained using ES rate directly with CC and NC rates \cite{bcc}  
\be
f_{B}=R^{CC}+\frac{(R^{NC}-R^{CC})(r-\bar{r}_{x})-(R^{ES}-R^{CC})(1-\bar{r}_{d})}{\bar{r}_{d}(r-\bar{r}_{x})-\bar{r}_{x}(1-\bar{r}_{d})}
\ee

We calculate the Borexino ES rate, $ R^{ES}_{Bor}$, once we know the survival probability for medium energy neutrinos, $P^{M}_{ ee}$. In order to determine the survival probability, we compare the Homestake event rate \cite{cl} with the SNO CC result. Since the fractional contributions of high energy $^{8}B$ and the medium energy neutrinos to 
the $^{37}Cl$ signals are about $80\%$ and $20\%$, respectively \cite{Barger} i.e.
\be
R_{Cl}=0.803R^{CC}_{SNO}+0.197P_{ee}^{M}.
\ee

The measured rate divided by the SSM prediction for the Homestake experiment gives the medium energy solar neutrino survival probability as 
\be
P^{M}_{ee}=\frac{R_{Cl}-0.803R^{CC}_{SNO}}{0.197} 
\ee

Using ES rate equation and the survival probability $P^{M}_{ee}$ for intermediate energy neutrinos, we obtain
\be
R^{ES}=P^{M}_{ee}+(1-P^{M}_{ee})[r\sin^{2}\alpha\sin^{2}\psi+\bar{r}_{x}\sin^{2}\alpha\cos^{2}\psi]
\ee 
where $r$ and $\bar{r}_{x}$ are cross-sectional ratios as defined earlier.

From the above equation, rate for no antineutrino component ($\sin^{2}\psi=1$) and a sterile admixture with the active neutrinos is given by
\be
R^{ES}_{Bor}=P_{ee}^{M}+ (1-P_{ee}^{M})r\sin^{2}\alpha
\ee
In a similar way, we can write expression for KamLAND Solar phase rate  with different value of $r$ \cite{thesiskamLand}. 
\section{Data Analysis}

In our analysis we use the Standard Solar Model predictions for $^{8}B$ solar neutrino flux $\phi_{SSM}=5.88 ^{+}_{-} 0.65 \times 10^{6}\ cm^{-2}s^{-1}$ expected to be detected in SNO and SK experiments \cite{SSM}.

For Homestake Chlorine experiment we use SSM prediction $8.09^{+}_{-} 1.09$ which corresponds to the rate  $R_{Cl}=0.32^{+}_{-}0.03$ \cite{bcc2:ga07}. Another rate leading to some interesting results is $R_{Cl}=0.337^{+}_{-}0.03$ \cite{Barger:2001pf}.  
The threshold energies are $E_{e_{th}}=5.5~MeV$, $5~MeV$, $7~MeV$, $5~MeV$, $0.665KeV$ \cite{0805} and $0.862MeV$ for SNO-II, SNO-III, SK-II, SK-III, Borexino and KamLAND respectively and the rest of the notation is standard. The cross sectional ratios are given in Table 3. The minor differences in the  values of $r,~\bar{r}_{x}~and~r_{d}$ pertaining to different experiments are mainly due to the difference in the threshold energies and are almost  independent of the resolution functions.
\begin{table}
\begin{small}
\begin{center}
\begin{tabular}{|c|c|c|c|}  \hline  
        &  $ r$                             & $\bar{r}_{x}$  &  $\bar{r}_{d}$        \\  \hline 
SNO-II  &  $0.150$            &  $0.115$    &  $0.954$        \\ 
SNO-III & $0.151$             & $0.116$     & $0.955$          \\  
SK-II   & $0.149$             & $0.114$     &   -              \\
SK-III  & $0.151$             & $0.116$     &   -               \\ 
Borexino& $0.213$             & $0.181$     &   -                \\
KamLAND & $0.210$             &    -        &   -                \\ \hline 
\end{tabular}
\end{center}
\caption{Cross sectional ratios for the SNO, SK, Borexino and KamLAND experiments.}
\end{small}
\end{table} 
Combining statistical and systematic errors, the second phase of SNO running (SNO-II) reports the three fluxes for a data set of 254.2 days and a 5.5 MeV energy threshold \cite{sno} as shown in Table 1. 
Table 1 also shows the data of 385.2 days by the third phase of SNO (SNO-III) following the installation of the NCDs with a 5 MeV energy threshold \cite{ncd}.
The Table 1 also includes the data of SK-II and SK-III fluxes \cite{sk}.

We divide our analysis into two cases. In Case-I we use fluxes $\phi^{NC},~\phi^{CC}$ from SNO-II with $\phi^{ES}$ from SNO-II, SK-II and SK-III one by one and in Case-II we use fluxes $\phi^{NC},~\phi^{CC}$ from SNO-III with $\phi^{ES}$ from SNO-III, SK-II and SK-III one by one and derive constraints on $f_{B}$, active, sterile neutrino and non-electronic neutrino and antineutrino fluxes.
\subsection{Constraints on active and sterile neutrinos}

We use the equations (3), (7) and (11) and we get the results as shown in Table 4. 

It is noted that, for no-sterile neutrinos, the CC/NC flux ratio in SNO is a direct measure of the average survival probability of $^{8}B$ solar neutrinos that were detected experimentally as 
$P_{ee}=\phi^{CC}/\phi^{NC}$. Solving it with errors, we will have $P_{ee}=0.340 ^{+}_{-} 0.04$ and $P_{ee}=0.301 ^{+}_{-} 0.03$, respectively for Case-I and Case-II. However, if we use SSM flux \cite{SSM} in place of NC flux then, $P_{ee}=\phi^{CC}/\phi_{SSM}$. $P_{ee}=0.286 ^{+}_{-} 0.04$ and $P_{ee}=0.284 ^{+}_{-} 0.04$ for Case-I and Case-II respectively. 
\begin{table}
\begin{small}
\begin{center}
\begin{tabular}{|c|c|c|c|c|c|c|}  \hline  
     & $\phi_{active}$ & $\phi_{sterile}$ & $\sin^{2}\alpha$ &$\phi_{no\bar{\nu}_{x}}^{ES}-\phi^{ES}$&$\gamma$& $\sin^{2}\psi$ \\  \hline 
&&&  {\bf Case-I}&&& \\ \hline
SNO-II &$4.64^{+}_{-}0.47$&$1.24^{+}_{-}0.80$ &$0.71^{+}_{-}0.16$ &$-0.18^{+}_{-}0.29$ &$0.21^{+}_{-}0.09$&$3.17^{+}_{-}3.54$              \\
SK-II&$4.59^{+}_{-}0.32$ & $1.29^{+}_{-}0.72$&$0.69^{+}_{-}0.13$&$-0.21^{+}_{-}0.20$
&$0.21^{+}_{-}0.07$&$3.62^{+}_{-}2.65$             \\ 
SK-III& $4.70 ^{+}_{-} 0.18$ & $1.18^{+}_{-}0.68$ &$0.72^{+}_{-}0.12$&$-0.15 ^{+}_{-} 0.12 $  &$0.20^{+}_{-}0.05$&$2.75^{+}_{-}1.73$               \\ \hline
&&& {\bf Case-II} &&&\\ \hline 
SNO-III &$6.31^{+}_{-}0.43$  &$-0.43^{+}_{-}0.78$ &$1.10^{+}_{-}0.20$&$-0.48^{+}_{-}0.27$ &$0.03^{+}_{-}0.07$&$-2.70^{+}_{-}2.02$              \\
SK-II &$5.33^{+}_{-}0.30$   & $0.55^{+}_{-}0.71$&$0.87^{+}_{-}0.15$&$-0.13^{+}_{-}0.19$  &$0.18^{+}_{-}0.06$&$2.29^{+}_{-}1.98$             \\ 
SK-III & $5.44 ^{+}_{-} 0.16$ & $0.44^{+}_{-}0.67$&$0.89^{+}_{-}0.15$&$-0.07 ^{+}_{-} 0.10 $&$0.17^{+}_{-}0.04$&$1.62^{+}_{-}1.23$               \\ \hline
\end{tabular}
\end{center}
\caption{Constraints on active, sterile and antineutrino fluxes for Case-I and Case-II in units of $10^{6}\ cm^{-2}s^{-1}$.}
\end{small}
\end{table}
The difference in above two results for NC and SSM flux indicates the possibility for sterile neutrino flux present in the solar neutrino flux.

In Table 4 column 1 and 2 show $\phi_{active}$ and $\phi_{sterile}$ at $1\sigma$ as obtained from equation (10) and (13) respectively. The $1\sigma$ upper bounds for the $\phi_{sterile}$ are obtained for the two cases as

\underline{Case-I}

$\phi_{sterile} \leq 2.04~\times 10^{6}~cm^{-2}~s^{-1}$,  \\
\hspace*{2cm}$\leq 2.01~\times 10^{6}~cm^{-2}~s^{-1}$,     \\
\hspace*{2cm}$\leq 1.86~\times 10^{6}~cm^{-2}~s^{-1}$,    \\

\underline{Case-II}

$\phi_{sterile} \leq 0.35~\times 10^{6}~cm^{-2}~s^{-1}$,  \\
\hspace*{2cm}$\leq 1.26~\times 10^{6}~cm^{-2}~s^{-1}$,         \\
\hspace*{2cm}$\leq 1.11~\times 10^{6}~cm^{-2}~s^{-1}$.         

It may be noted that these bounds are more constrained as obtained earlier in literature \cite{skumar}. We found a possibility of no-sterile solar neutrino flux in the lower side of $1\sigma$ in Case-II.

The column 3 of Table 4 represent the percentage of active neutrinos present in the solar neutrino flux. The $1\sigma$ range of $\sin^{2}\alpha$ for Case-I indicate strong possibility, i.e. up to $20\%$, of sterile neutrino fraction (which is proportional to $\cos^{2}\alpha$). However, in the Case-II there is a possibility for no-sterile fraction at  $1\sigma$.
\subsection{Constraints on non-electronic neutrinos and antineutrinos}
 
The $\phi_{\nu_{x}}$ as obtained from equation (11) and (12) is shown in Table 5. The NC flux is consistent with SSM predictions showing that the SSM is correctly modelling the Sun. The flux deficit in the reactions sensitive to electronic flavour neutrino only shows possibility of flavour change. This effect can be understood by plotting the results of three different reactions of neutrinos detection i.e. $\phi^{CC},~\phi^{NC}~and~\phi^{ES}$ reactions against $\phi^{CC}$. 
It may be noted that the ES reaction gives a diagonal band with a slope more sensitive to electronic flavour neutrinos than the NC reaction. On the other hand, the CC reaction gives a vertical band as it is sensitive to electronic flavour neutrinos only. These measurements overlap as shown in the Figure 1 indicating presence of neutrino flavour change.
\begin{figure}
\begin{center}
\epsfig{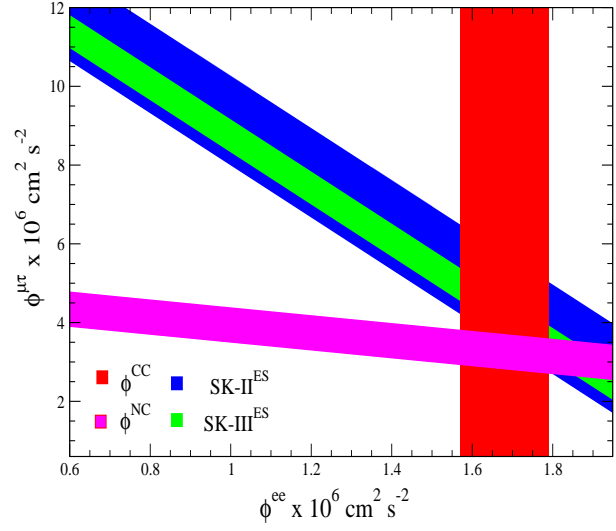}\\
\vspace{3cm}
\epsfig{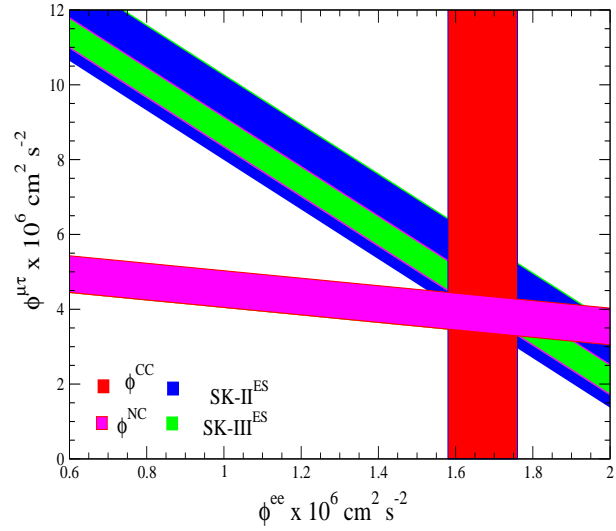}
\end{center}
\caption{\ $\phi^{CC}$~ vs~ $\phi_{\mu \tau}$~ flux for Case-I and Case-II}
\end{figure}

The column 4, 5 and 6 of Table 4 are the parameters associated with non-electronic neutrinos and antineutrinos. The positive sign of the difference $\phi^{ES}_{no \bar{\nu_{x}}}-\phi^{ES}$ in Column 4 will indicate the presence of non-electronic antineutrinos ($\bar{\nu_{x}}$) in the solar neutrino flux. The numerical value of $\phi_{\bar{\nu_{x}}}$ component is obtained from equation (9). 

The presence of $\phi_{\bar{\nu_{x}}}$ is disallowed for Case-I (SK-II and SK-III ES) and for Case-II (SNO-III ES) at $1\sigma$. However, $\phi_{\bar{\nu_x}}$ is present for all other sub-cases. So the corresponding upper bounds at $1\sigma$ are as follows:

\begin{table}
\begin{small}
\begin{center}
\begin{tabular}{|c|c|c|c|}  \hline  
   &     $\phi^{ES}_{\nu_x}$   SK-II     & $\phi^{ES}_{\nu_{x}}$ SK-III  &  $\phi^{NC}_{\nu_{x}}$  \\ \hline
& & &  \\ 
Case-I & $4.47^{+}_{-}1.94$  & $4.70^{+}_{-}1.36$  &$3.26^{+}_{-}0.44$                \\
Case-II&  $4.77 ^{+}_{-} 1.29$ & $4.30 ^{+}_{-} 0.72$& $3.87^{+}_{-}0.50$                  \\ \hline 
\end{tabular}
\end{center}
\caption{Predictions for $\phi_{\nu_{x}}$ for Case-I and Case-II in units of $10^{6}\ cm^{-2}s^{-1}$}.
\end{small}
\end{table}
$\phi_{\bar{\nu_{x}}} \leq ~ 3.78~\times 10^{6}~cm^{-2}~s^{-1}$ Case-I (SNO-II ES) \\
\hspace*{1.5cm}$\leq~ 2.00~\times 10^{6}~cm^{-2}~s^{-1}$  Case-II (SK-II ES)  \\
\hspace*{1.5cm}$\leq~ 0.75~\times 10^{6}~cm^{-2}~s^{-1}$  Case-II (SK-III ES).

We do not obtain any inference about the $\nu_x ,\bar{\nu}_{x}$ admixture in the solar neutrino flux for Case-II (SNO-III ES), however, the Case-II (SK-II and SK-III ES) shows presence of non-electronic solar neutrino flux,   which is also evident from the bounds obtained as above for $\phi_{\bar{\nu}_{x}}$ at $1 \sigma$.
The results obtained above are in agreement with the one presented in the literature  \cite{skumar} where the maximum possible antineutrino flux limit has been mentioned. 

As stated earlier in equation (14), the fraction of non-electronic neutrinos in $\nu_{x},\bar{\nu_{x}}$ admixture present in the solar neutrino flux is given by $\sin^{2}\psi$. $\sin^{2}\psi$ for Case-I with SNO-II ES gives arbitrary $\nu_{x},\bar{\nu_{x}}$ admixture but for SK-II and SK-III ES, i.e. the second and third rows, indicate the minimum antineutrino flux component in the solar neutrinos. 
\subsection{Constraints on $f_{B}$}
We use the values of $\sin^{2}\psi$ and $\sin^{2}\alpha$ from equations (15) and (16) respectively and calculate $f_{B}$ using equations (20) and (21) for Case-I and Case-II. The results are shown in Table 6. Since no inference on $\sin^{2}\psi$ for Case-II (SNO-III ES) can be obtained as shown in Table 4, we compute the constraints on $f_{B}$ only for Case-II (SK ES). The central value of $f_B$ for Case-I as well as for Case-II is greater than one for the so called `sterile case' and decreases as $\sin^{2}\psi$ approaches unity. However, for no-sterile case the central value $f_B$ has been found to be less than one and follow the same variation as in sterile case.
We have also calculated $f_B$ for no-sterile case from equation (22) using ES rates directly as shown in Table 7.
\begin{table}[h]
\begin{small}
\begin{center}
\begin{tabular}{|c|c|c|c|}  \hline  
ES $\rightarrow$  &SNO-II                           & SK-II                                                                 &  SK-III  \\
        &$\sin^{2}\psi=1$      & $\sin^{2}\psi=1$ \ $\sin^{2}\psi=0.97$                      & $\sin^{2}\psi=1$  \\  \hline 
$f_{B}$(sterile case) & $1.07^{+}_{-}0.21$  &$1.09^{+}_{-}0.19$ \ $1.09^{+}_{-}0.19$  &$1.06^{+}_{-}0.17$                \\
$f_{B}$(no-sterile case)&  $0.84 ^{+}_{-} 0.07 $ & $0.84 ^{+}_{-} 0.07$\ $0.84^{+}_{-}0.08$ & $0.84^{+}_{-}0.07$                  \\ \hline
ES $\rightarrow$     &SNO-III                           & SK-II            &  SK-III  \\
        &     -      & $\sin^{2}\psi=1$ \ $\sin^{2}\psi=0.31$                      & $\sin^{2}\psi=1$ \ $\sin^{2}\psi=0.39$  \\  \hline 
&&& \\
$f_{B}$(sterile case) & -  &$1.04^{+}_{-}0.16$ \ $1.06^{+}_{-}0.17$  &$1.02^{+}_{-}0.16$\  $1.04^{+}_{-}0.16$  \\
$f_{B}$(no-sterile case)&  - & $0.94 ^{+}_{-} 0.08$\ $0.96^{+}_{-}0.09$ & $0.94^{+}_{-}0.08$\ $0.96^{+}_{-}0.09 $\\\hline 
\end{tabular}
\end{center}
\caption{Predictions for $f_{B}$ for Case-I and Case-II}
\end{small}
\end{table}
\begin{table}[h]
\begin{small}
\begin{center}
\begin{tabular}{|c|c|c|c|}  \hline  
ES $\rightarrow$        &      SNO                       & SK-II                            &  SK-III  \\   \hline 
$f_{B}$(no-sterile case) &                                  &                    &                    \\
 (Case-I)                       & $0.79^{+}_{-}0.13$ &$0.78^{+}_{-}0.11$   &$0.80^{+}_{-}0.10$\\
$f_{B}$(no-sterile case) &           &                                     &                     \\
(Case-II)                       & $1.07^{+}_{-}0.13$ & $0.91 ^{+}_{-}0.12$ & $0.92^{+}_{-}0.11 $ \\ \hline 
\end{tabular}
\end{center}
\caption{Predictions for $f_{B}$ using ES rates directly (for no-sterile case)}
\end{small}
\end{table}

$f_B$ is consistent with the results obtained earlier by Chauhan $\it {et~ al}.$ \cite{bcc}. The slight difference is due to the fact that we have taken SK-II and SK-III data along with SNO-II and SNO-III which has slight difference from earlier data. The variation of $f_{B}$ with $\sin^{2}\alpha$ for the lower and upper bounds on $\sin^{2}\psi$ obtained from the analysis is shown in the Figure 2 and Figure 3 for Case-I and Case-II respectively.
\begin{figure}
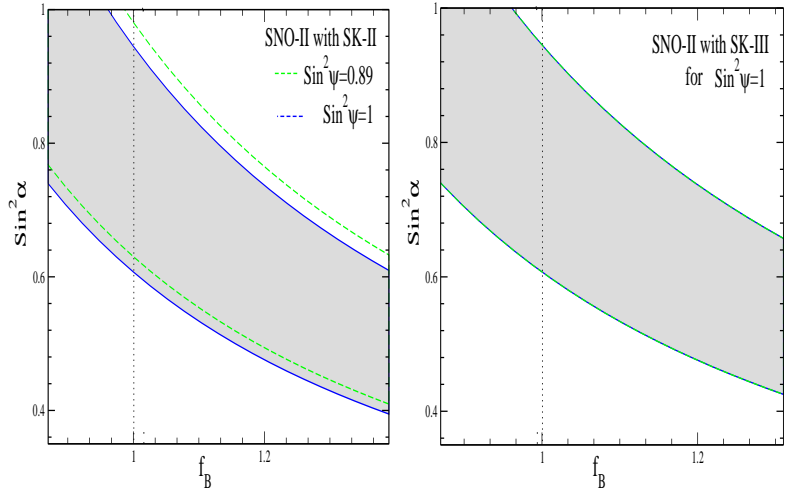

\begin{center} 
\epsfig{file=2ab.eps,height=6.5cm,width=5.1cm}
\epsfig{file=3ab.eps,height=6.5cm,width=5.1cm}
\end{center}
\caption{$f_{B}-\sin^{2}\alpha$ degeneracy for Case-I}
\end{figure}
\begin{figure}
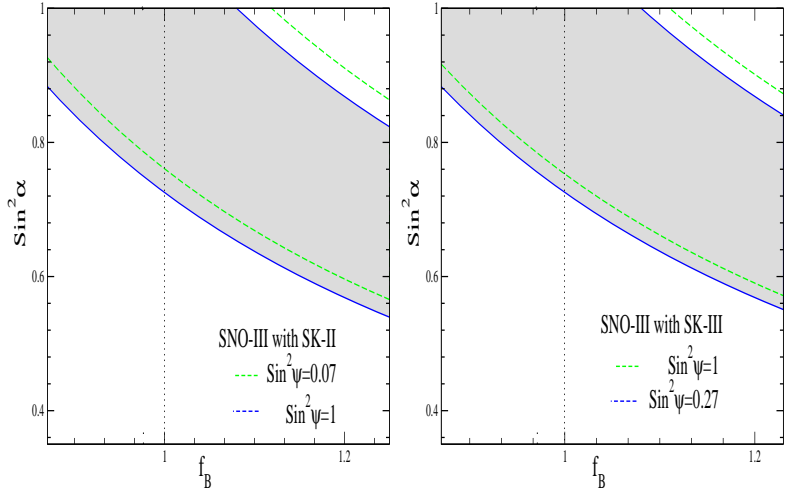

\begin{center}
\epsfig{file=2cd.eps,height=6.5cm,width=5.1cm}
\epsfig{file=3cd.eps,height=6.5cm,width=5.1cm}
\end{center}
\caption{$f_{B}-\sin^{2}\alpha$ degeneracy for Case-II}
\end{figure} 
\subsection{Predictions for Borexino and KamLAND}
Using the model independent rates equations (25)and (26), we predict rates for Borexino and KamLAND intermediate energy solar neutrinos. We have the following results:
 
Using $R_{Cl}=0.32^{+}_{-} 0.03$ \cite{bcc2:ga07} and $R^{CC}$, we get the numerical
value
\begin{equation}
P^{M}_{ee}=0.441^{+}_{-} 0.17~and~P^{M}_{ee}=0.449^{+}_{-} 0.17
\end{equation}
for Case-I and Case-II respectively. The rates predicted for Borexino and KamLAND are shown in Tables 8 and 9, respectively.
\begin{table}[h]
\begin{small}
\begin{center}
\begin{tabular}{|c|c|c|c|}  \hline  
ES $\rightarrow$        & SNO-II                           & SK-II                                                                 &  SK-III  \\
        &$\sin^{2}\psi=1$      & $\sin^{2}\psi=1$ \ $\sin^{2}\psi=0.97$                      & $\sin^{2}\psi=1$   \\  \hline 
&&& \\        
$R^{ES}_{Bor}$(sterile case) & $0.525^{+}_{-}0.18$  &$0.523^{+}_{-}0.18$ \ $0.522^{+}_{-}0.18$  &$0.526^{+}_{-}0.18$             \\
$R^{ES}_{Bor}$(no-sterile case)&  $0.560 ^{+}_{-} 0.17 $ & $0.560 ^{+}_{-} 0.17$\ $0.559^{+}_{-}0.17$ & $0.560^{+}_{-}0.17$   \\ \hline
ES $ \rightarrow$      &SNO-III                           & SK-II            &  SK-III  \\
        &        -      & $\sin^{2}\psi=1$ \ $\sin^{2}\psi=0.31$                      & $\sin^{2}\psi=1$ \ $\sin^{2}\psi=0.39$  \\  \hline 
&&& \\
$R^{ES}_{Bor}$(sterile case)  &   -   &$0.551^{+}_{-}0.18$ \ $0.540^{+}_{-}0.18$  &$0.553^{+}_{-}0.18$\  $0.544^{+}_{-}0.18$\\
$R^{ES}_{Bor}$(no-sterile case)& - & $0.566 ^{+}_{-} 0.17$\ $0.554^{+}_{-}0.17$ & $0.566^{+}_{-}0.17$\ $0.555^{+}_{-}0.17 $  \\ \hline
\end{tabular}
\end{center}
\caption{Predictions for $R^{ES}_{Bor}$ for Case-I and Case-II}
\end{small}
\end{table} 
\begin{table}[h]
\begin{small}
\begin{center}
\begin{tabular}{|c|c|c|c|}  \hline  
ES $ \rightarrow$      &SNO-II                           & SK-II            &  SK-III  \\
        &$\sin^{2}\psi=1$      & $\sin^{2}\psi=1$& $\sin^{2}\psi=1$ \\  \hline 
&&& \\
$R^{ES}_{KamL}$(sterile case) & $0.524^{+}_{-}0.18$  &$0.522^{+}_{-}0.18$ &$0.525^{+}_{-}0.18$        \\
$R^{ES}_{KamL}$(no-sterile case)&  $0.558 ^{+}_{-}0.17 $ & $0.558 ^{+}_{-} 0.17$  & $0.558^{+}_{-}0.17$  \\ \hline
ES $ \rightarrow$      &SNO-III                           & SK-II            &  SK-III  \\ \hline 
&&& \\
$R^{ES}_{KamL}$(sterile case) &  - &$0.549^{+}_{-}0.18$&$0.552^{+}_{-}0.18$     \\
$R^{ES}_{KamL}$(no-sterile case)& - & $0.564 ^{+}_{-} 0.17$& $0.564^{+}_{-}0.17$ \\ \hline 
\end{tabular}
\end{center}
\caption{Predictions for $R^{ES}_{KamL}$ for Case-I and Case-II}
\end{small}
\end{table}

However, it is interesting to note that if we use $R_{Cl}=0.337^{+}_{-}0.03$ \cite{Barger:2001pf}, we obtain
\be
P^{M}_{ee}=0.545^{+}_{-}0.17~and~ P^{M}_{ee}=0.553^{+}_{-}0.17
\ee
for Case-I and Case-II, respectively having central values consistent as given in \cite{Marco}. The difference in $P^{M}_{ee}$ in equations (27) and (28) is due to the different values predicted by the SSM for the Homestake experiment. However, experimental result i.e. $R_{Cl}$ being the same. Using equations (25) and (26) we predict the rates as given in the Tables 10 and 11 for Borexino and KamLAND experiments, respectively, in a model independent way.
\begin{table}[h]
\begin{small}
\begin{center}
\begin{tabular}{|c|c|c|c|}  \hline  
ES $\rightarrow$        &SNO-II                           & SK-II                                                                 &  SK-III  \\
        &$\sin^{2}\psi=1$      & $\sin^{2}\psi=1$ \ $\sin^{2}\psi=0.97$                      & $\sin^{2}\psi=1$   \\  \hline 
&&& \\        
$R^{ES}_{Bor}$(sterile case) & $0.621^{+}_{-}0.18$  &$0.619^{+}_{-}0.18$ \ $0.618^{+}_{-}0.18$  &$0.622^{+}_{-}0.18$             \\
$R^{ES}_{Bor}$(no-sterile case)&  $0.648 ^{+}_{-} 0.18 $ & $0.648 ^{+}_{-} 0.18 $\ $0.648 ^{+}_{-} 0.18 $ & $0.648 ^{+}_{-} 0.18 $   \\ \hline
ES $ \rightarrow$      &SNO-III                           & SK-II            &  SK-III  \\
        &        -      & $\sin^{2}\psi=1$ \ $\sin^{2}\psi=0.31$                      & $\sin^{2}\psi=1$ \ $\sin^{2}\psi=0.39$  \\  \hline
&&& \\         
$R^{ES}_{Bor}$(sterile case) & -  &$0.636^{+}_{-}0.19$ \  $0.627^{+}_{-}0.19$  &$0.638^{+}_{-}0.19$\  $0.630^{+}_{-}0.19$\\
$R^{ES}_{Bor}$(no-sterile case)&  - & $0.648 ^{+}_{-} 0.18$\  $0.648 ^{+}_{-} 0.18$ & $0.648 ^{+}_{-} 0.18$\ $0.648 ^{+}_{-} 0.18$  \\ \hline
\end{tabular}
\end{center}
\caption{Predictions for $R^{ES}_{Bor}$ for Case-I and Case-II}
\end{small}
\end{table}
\begin{table}[h]
\begin{small}
\begin{center}
\begin{tabular}{|c|c|c|c|}  \hline  
ES $ \rightarrow$      &SNO-II                           & SK-II         &  SK-III  \\
        &$\sin^{2}\psi=1$      & $\sin^{2}\psi=1$                       & $\sin^{2}\psi=1$   \\  \hline 
&&& \\
$R^{ES}_{KamL}$(sterile case) & $0.613^{+}_{-}0.18$  &$0.610^{+}_{-}0.18$   &$0.614^{+}_{-}0.18$        \\
$R^{ES}_{KamL}$(no-sterile case)&  $0.640 ^{+}_{-}0.18$ & $0.640 ^{+}_{-}0.18$ & $0.640 ^{+}_{-}0.18$  \\ \hline
ES $ \rightarrow$      &SNO-III                           & SK-II         &  SK-III  \\  \hline
&&& \\
$R^{ES}_{KamL}$(sterile case) & -&$0.635^{+}_{-}0.19$ &$0.637^{+}_{-}0.19$ \\
$R^{ES}_{KamL}$(no-sterile case)& - & $0.640 ^{+}_{-} 0.18$ & $0.640 ^{+}_{-} 0.18$   \\ \hline
\end{tabular}
\end{center}
\caption{Predictions for $R^{ES}_{KamL}$ for Case-I and Case-II}
\end{small}
\end{table}
\subsection{\textbf{$\chi^{2}$-Fitting }}
     
In this subsection we perform the $\chi^{2}$ analysis for the Case-I and Case-II. The $\chi^{2}$ definition used is as follows
\be
\chi^{2}=\sum_{i}\frac{(R_{i}-R^{th}_{i})^{2}}{\delta R^{2}_{i}}.
\ee
The sum is extended over the five experiments($i=ES_{SK},\ ES_{SNO},\ NC,\ CC.$) where $R_{i}$ and $\delta R_{i}$ denote the experimental rates and their errors as quoted in Table 2 and $R_{i}^{th}$ are given by equations (17)-(19).
\begin{table}[h]
\begin{small}
\begin{center}
\begin{tabular}{|c|c|c|c|c|c|}  \hline  
  &$f_{B}$&$P_{ee}$&$\sin^{2}\alpha$&$\sin^{2}\psi$&$\chi^{2}_{min}$           \\   \hline                                           
&&&&& \\
LMA(3 dof)    &0.883    &0.346    &1    &1       &1.86                 \\
LMA+$\bar{\nu}_{x}$(2 dof) &0.883   &0.346 &1     &1    &1.86 \\
LMA+$\nu_{s}$(2 dof)     &0.917       &0.334  &0.94   &1 &1.86   \\
LMA+$\bar{\nu}_{x}+\nu_{s}$(1 dof)&0.890&0.343&0.98&1&1.86 \\ \hline
\end{tabular}
\end{center}
\caption{Best fit values by $\chi^{2}$- Analysis for Case-I}
\end{small}
\end{table}
\begin{table}[h]
\begin{small}
\begin{center}
\begin{tabular}{|c|c|c|c|c|c|}  \hline  
  &$f_{B}$&$P_{ee}$&$\sin^{2}\alpha$&$\sin^{2}\psi$&$\chi^{2}_{min}$           \\ \hline                                              
&&&&& \\
LMA(3 dof)   &0.956   &0.304  &1    &1       &5.95           \\
LMA+$\bar{\nu}_{x}$(2 dof)&0.956&0.304&1&1&5.95  \\
LMA+$\nu_{s}$(2 dof)&1.005&0.289&0.932&1&5.95          \\
LMA+$\bar{\nu}_{x}+\nu_{s}$(1 dof)&1.000&0.291&0.944&1&5.95  \\ \hline
\end{tabular}
\end{center}
\caption{Best fit values by $\chi^{2}$- Analysis for Case-II}
\end{small}
\end{table}

An inspection of Table 12 and Table 13 shows that the best fit for LMA+$\bar{\nu}_{x}$ corresponds to the very absence of antineutrino component ($\bar{\nu}_{x}$) i.e. $\sin^{2}\psi=1$. It is also seen that allowing for $\nu_s$ alone in addition to LMA ($LMA+\nu_{s}$) as well as $LMA+\bar{\nu}_{x}+\nu_{s}$ lead to a best fit solution with a small $\nu_s$ component (upto $7\%$). The best fit value for $\sin^{2}\psi$ is independent of $\nu_{s}$ component and of the parameters $f_{B},~P_{ee}$.
\section{Conclusions}
In the present work, we have derived in a model independent way the constraints on $\nu_s$, $\bar{\nu_{x}}$ and 
the SSM normalization factor $f_{B}$ using the recent data of SNO and SK experiments. We calculated the upper bound on the flux of  $\phi_{sterile}$ and $\phi_{\bar{\nu_{x}}}$ solar neutrinos. The medium energy survival probability has been calculated for Borexino and compared with the existing results \cite{Marco}, the medium energy survival probability and rates for KamLAND has been predicted. The hints for the possible solar neutrinos has emerged from WMAP data as well which suggest that the number of neutrino families in the early universe was four \cite{wmap}. The antineutrino data when combined with other experimental data eads to the possibility of a 3+1 model i.e. the three ordinary neutrino and a sterile one \cite{kara}. The experiments like MINOS \cite{minos} are running and searching for the existence of sterile neutrinos. The results of this work can be summarized as follows:- 
\begin{enumerate} 
\item[(i)]  Non-electronic antineutrino component strictly disallowed in Case-I (SK-II and SK-III ES) at $1\sigma $
\item[(ii)] Non-electronic antineutrino allowed for Case-I (SNO-II ES) and Case-II for (SK-II and SK-III ES) at $1\sigma$.
\item[(iii)] Active solar neutrino flux upper and lower bounds in the sub-case of Case-I (SK-II and SK-III ES) exist i.e. clearly suggesting the presence of sterile neutrino components in solar neutrino flux. However, as suggested by Case-II (SK-II and SK-III ES) this may/may not be present in the solar neutrino data.
\item[(iv)] The upper bounds on sterile flux are more constrained than previously obtained in literature \cite{skumar}. 
\item[(v)] Bounds on maximum possible non-electronic neutrino component has been obtained.
\item[(vi)] Borexino rates has been calculated and compared with the existing data and predictions for KamLAND rate has been made. 
\item[(vii)] $\sin^{2}\psi=1$ i.e. no antineutrino flux possibility is most favourable and sterile neutrinos are allowed upto $7\%$ are suggested by the $\chi^2$-Fitting.   
\end{enumerate}

\section*{Acknowledgements}
The research work of S.D. and L.S. is supported by the University Grants Commission, Government of India {\it vide} Grant No. 34-32/2008~(SR). One of the authors B.C.C. thanks IUCAA for the providing hospitality during the preparation of the work. 

\end{document}